\documentclass{ws-mplb}

\usepackage{}

\begin{document}

%\markboth{Authors' Names}{Instructions for
%Typing Manuscripts (Paper's Title)}

\catchline{}{}{}{}{}
%
%%%%%%%%%%%%%%%%%%%%%%%%%%%%%%%%%%%%%%%%%%%%%%%%%%%%%%%%%%%%%%%%%%%%

\title{Multiple Superconducting Gaps, Anisotropic Spin Fluctuations and Spin-Orbit Coupling in Iron-Pnictides}

\author{Jie Yang\footnote{yangjie@iphy.ac.cn}}

\address{Institute of Physics, Chinese Academy of Sciences\\Beijing, 100190, P.R.China}

\author{Guo-qing Zheng}

\address{Institute of Physics, Chinese Academy of Sciences\\Beijing, 100190, P.R.China\\
Department of Physics, Okayama University\\Okayama, 700-8530, Japan}

\maketitle

\begin{history}
\received{(26 April 2012)}
%\revised{(Day Month Year)}
\end{history}

\begin{abstract}
This article reviews the NMR and NQR studies on iron-based high-temperature superconductors by the IOP/Okayama group. It was found that the electron pairs in the superconducting state are in the spin-singlet state with multiple fully-opened energy gaps. The antiferromagnetic spin fluctuations in the normal state are found to be closely correlated with the superconductivity. Also the antiferromagnetic spin fluctuations are anisotropic in the spin space, which is different from the case in copper oxide superconductors. This anisotropy originates from the spin-orbit coupling and is an important reflection of the multiple-bands nature of this new class of superconductors.
\end{abstract}


\section{Introduction}

The copper oxide high-temperature superconductors discovered in 1986 opened a new chapter in the history of superconductivity research. The critical transition temperature ($T_{c}$) was quickly raised up to 135 K~\cite{Cu_135} at ambient pressure and 164 K~\cite{Cu_164} under high pressure. This record has held since 1994. During the past two decades, great efforts have been made to explore new types of high-$T_{c}$ superconductors. In particular, the success of the copper oxides naturally makes other transition metal oxides the promising candidates for new high-$T_{c}$ families. In 2003, Takada {\it et al.} found superconductivity in sodium cobalt oxide,~\cite{NaCoO_5K} but $T_{c}$ is only about 5 K. In 2008, a breaking was made by Kamihara {\it et al.} who reported superconductivity of 26 K in LaFeAsO$_{1-x}$F$_{x}$.~\cite{26K_hosono} Shortly after that, by replacing La with other rare-earth (RE) elements Ce, Pr, Nd, Sm and Gd, a series of fluorine-doped superconductors with $T_{c}$ higher than 50 K were fabricated~\cite{ce1111_41K,Pr52K,Nd1111,Sm43K,Gd53K}, and a record of $T_{c}$ = 55 K was made in SmFeAsO$_{1-x}$F$_{x}$~\cite{Sm55K} and Gd$_{1-x}$Th$_{x}$FeAsO~\cite{GdTh}. The record-holding compounds have the same crystal structure and are denoted as 1111 type iron-base superdonductors. Besides  of  the  1111  type, other structures were also synthesized, including hereafter AFe$_{2}$As$_{2}$~\cite{Ba122_38K} and A$_{x}$Fe$_{2}$Se$_{2}$~\cite{KFe2Se2} series (A = alkaline earth metal or alkali metal, abbreviated as 122 type), LiFeAs series~\cite{111_jin,111_chu}(abbreviated as 111 type), FeSe series~\cite{FeSe} (abbreviated as 11 type), and some compounds with multiple-layer structures~\cite{42622}. These iron-based superconductors have become the second class of high-$T_{c}$ family. They share some common features. (i) They all have a 2D layered structure, in which conducting Fe-As layers are separated by other blocks of insulating RE-O(F) or ionic A$^{2+}$. (ii) Elemental substitution or application of external pressure suppresses a structural and/or magnetic phase transition and leads to superconductivity.

This article reviews the Nuclear Magnetic Resonance (NMR) and Nuclear Quadrupole Resonance (NQR) studies on several iron-based high-$T_{c}$ superconductors carried out by the IOP/Okayama group. Some results on the related topics from other NMR groups are also included. For more comprehensive reviews, readers are referred to papers by Hirschfeld {\it et al.}~\cite{Hirschfeld} and Stewart~\cite{Stewart}. This paper is organized as follows: In Sec. 2, the superconducting properties of iron-pnictides are described. In Sec. 3, the relationship between antiferromagnetic spin fluctuations and superconductivity is discussed. Finally, the anisotropy of the antiferromagnetic spin fluctuations is discussed in Sec. 4.

\section{Superconducting properties}

The nuclear spin-lattice relaxation rate $1/T_{1}$ and the Knight shift $K$ measured by NMR/NQR are important data for determining the superconducting gap structure and the Cooper-pair spin symmetry. In this section, we will summarize the NMR/NQR experimental results on the superconducting properties of various iron-based superconductors.

\subsection{PrFeAsO$_{1-x}$F$_{x}$}

Matano {\it et al.} reported the results of Knight shift ($K$) and $1/T_{1}$ in PrFeAsO$_{0.89}$F$_{0.11}$ with $T_{c}$ = 45 K~\cite{matano_PrFeAsO}. The polycrystalline PrFeAsO$_{0.89}$F$_{0.11}$ sample was synthesized by the high-pressure method~\cite{Pr52K}. For NMR measurements, the sample was crushed into powders, then aligned in a magnetic field of $H$ = 9 T and fixed by epoxy.

\begin{figure}[th]
\centerline{\psfig
{file=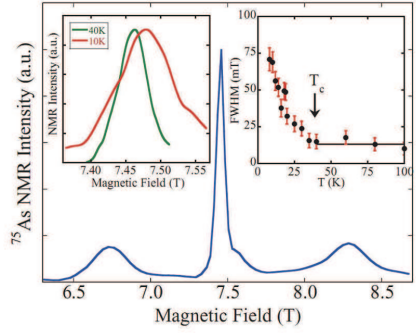,width=6cm}}
\vspace*{8pt}
\caption{(Color online) $^{75}$As-NMR spectrum of PrFeAsO$_{0.89}$F$_{0.11}$ with $H$~$\|$~$ab$-plane at 55.1 MHz and $T$ = 40 K. The right inset shows the full width at the half-maximum (FWHM) of the central transition peak as a function of temperature. The left inset compares the central transition peak at $T$ = 40 K.}
\label{f1}
\end{figure}

\begin{figure}[th]
\centerline{\psfig{file=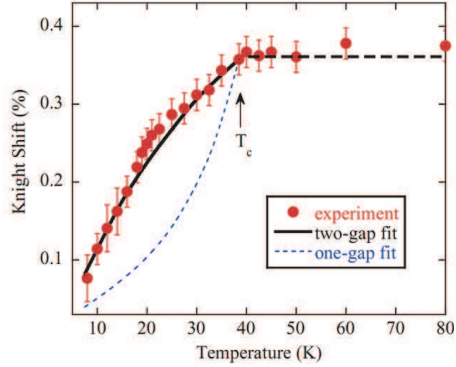,width=6cm}}
\vspace*{8pt}
\caption{(Color online) The temperature variation of the $^{75}$As Knight shift with $H$~$\|$~$ab$ in PrFeAsO$_{0.89}$F$_{0.11}$. The solid curve is a fitting of two gaps with $\Delta_{1}$($T$ = 0) = 3.5$k_{B}$$T_{c}$ and a relative weight of 0.4, and $\Delta_{2}$($T$ = 0) = 1.1$k_{B}$$T_{c}$ with a relative weight of 0.6. The broken curve below $T_{c}$ is a simulation for the larger gap
alone. In both cases, $K_{orb}$ was taken as 0.008\%.
}\label{f2}
\end{figure}

Figure~\ref{f1} shows a typical $^{75}$As NMR spectrum of PrFeAsO$_{0.89}$F$_{0.11}$ at 40 K, which consists of a sharp central peak and two satellite peaks due to the nuclear quadrupole interaction. The width of the peak is temperature independent above $T_{c}$, while it increases below $T_{c}$ due to the formation of a vortex lattice in the superconducting state, which confirms the bulk nature of the superconductivity. The Knight shift as a function of temperature is shown in Fig.~\ref{f2}. The decrease of $K$ to almost zero indicates the spin-singlet pairing. Another important feature is that,  $K$ decreases below $T_{c}$ down to $T$ = 20 K, then is followed by a still sharper drop below. Such behavior is not seen in usual superconductors such as copper oxides, where $K$ decreases rapidly below $T_{c}$ and is followed by a milder decrease at low temperatures, as illustrated by the broken curve in Fig.~\ref{f2}

\begin{figure}[th]
\centerline{\psfig{file=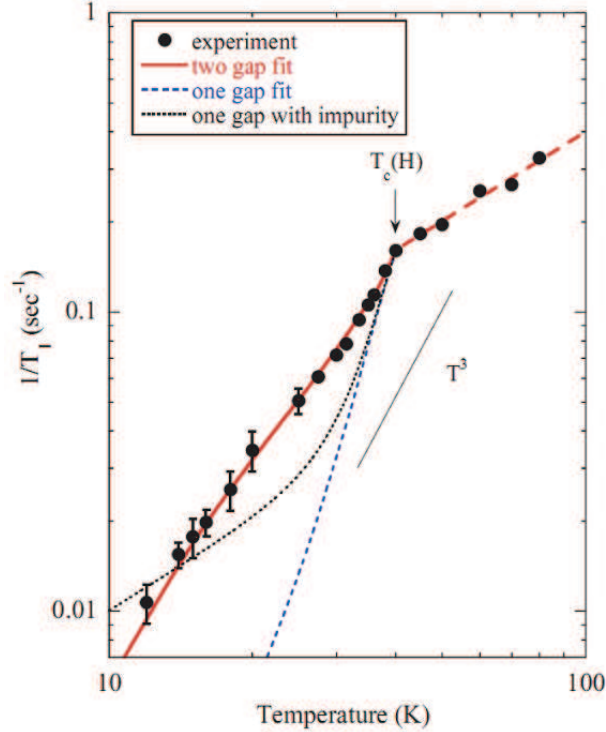,width=8cm}}
\vspace*{8pt}
\caption{(Color online) The temperature dependence of $^{19}$F spin-lattice relaxation rate $1/T_{1}$ in PrFeAsO$_{0.89}$F$_{0.11}$ measured at $H$ = 1.375 T. The solid curve is a two-gap fit with the same parameters as in Fig.~\ref{f2}. The broken curve below $T_{c}$ is a simulation for the larger gap alone, and the dotted curve is for the case of impurity scattering with 50\% DOS remained at the Fermi level. The thin straight line illustrates the temperature dependence of $T^{3}$.
}\label{f3}
\end{figure}

The step-wise decrease of the Knight shift is also reflected in the temperature dependence of the $^{19}$F spin-lattice relaxation rate $1/T_{1}$, as seen in Fig.~\ref{f3}. $1/T_{1}$ is in proportion to $T$ above $T_{c}$, but drops sharply below $T_{c}$. The $1/T_{1}$ shows no coherence peak just below $T_{c}$, which is not consistent with a conventional $s$-wave gap. For an isotropic $s$-wave fully-opened gap, $1/T_{1}$ would show a coherence peak just below $T_{c}$. Moreover, there is a broad hump-like feature around $T \sim T_{c}/2$. Such behavior is not expected in a single-gap superconductor. Matano {\it et al.} found that $1/T_{1}$  is of single component throughout the whole temperature range, and thus rule out the possibility of impurity as a possible cause for the uncommon $T$-variation of $1/T_{1}$.

Matano {\it et al.} suggested that a two-gap model can explain the step-wise temperature variation of both $K$ and $1/T_{1}$. The underlying physics is that the physical quantities just below $T_{c}$ are dominantly governed by a larger gap while the system does not ¡°notice¡± the existence of a smaller gap. Only at low temperatures where the thermal energy becomes comparable to or smaller than the smaller gap, does the system realize the smaller gap, resulting in another drop of $K$ and $1/T_{1}$. The spin-lattice relaxation rate in the superconducting state ($1/T_{1s}$) can be expressed as

{\footnotesize
\begin{equation}\label{e1}
 \frac{{{T_{1N}}}}{{{T_{1s}}}} = \sum\limits_i {\frac{2}{{{k_B}T}}\int{\int{\left( {1 + \frac{{{\Delta _i}^2}}{{EE'}}} \right){N_{s,i}}(E){N_{s,i}}(E')f(E)\left[{1 - f(E')}\right]\delta \left( {E - E'} \right)dEdE'} } }
\end{equation}
}

where $i$ denotes the number of different gap or band, $C = 1 + \frac{{{\Delta ^2}}}{{EE'}}$ is the coherence factor, ${{\Delta }_{i}}$ is the superconducting gap on band $i$, ${N_{s,i}}(E) = {N_{s,i}}\frac{E}{{\sqrt {{E^2} - {\Delta _i}^2} }}$ is the density of states(DOS) in the superconducting state. The $1/T_{1}$ data can be fitted well by using this two gap model as is shown in Fig.~\ref{f3}. In this early work, the authers assumed a $d$-wave gap. Later, they used a multiple-gap $s^{\pm}$-wave model to fit their data in LaFeAsO$_{1-x}$F$_{x}$, Ba$_{1-x}$K$_{x}$Fe$_{2}$As$_{2}$ and LiFeAs (see below).

This is the first work pointing out the multiple-gap feature of iron-based superconductors, and linking the multiple gaps to the multiple-band structure of this class of materials. Band calculation indicates that the Fermi surface (FS) of LaFeAsO consists of hole pockets around the $\Gamma$ point and electron pockets around the
$M$ point~\cite{singh,fang}. Soon after Matano's paper, Ding {\it et al.} used the angle-resolved photoemission spectroscopy (ARPES) technique and found three gaps in Ba$_{1-x}$K$_{x}$Fe$_{2}$As$_{2}$~\cite{ARPES_ding}.

\subsection{LaFeAsO$_{1-x}$F$_{x}$}
\label{LaFeAsO(F)}

The two-gap feature was also found in LaFeAsO$_{0.92}$F$_{0.08}$ ($T_{c}$ = 23 K) by Kawasaki {\it et al.}~\cite{LaFeAsO_Kawasaki}, who suggested that either $d$-wave or $s^{\pm}$-wave model could count for their data. Earlier, Grafe {\it et al.}~\cite{LaFeAsO_grafe} reported a $1/T_{1} \sim T^{3}$ behavior, and considered as evidences for line nodes. Nakai {\it et al.} reported the same behavior for LaFeAsO$_{1-x}$F$_{x}$~\cite{LaFeAsO_nakai}, and found the $1/T_{1}$ shows robust $T^{3}$ behavior in the field range from 5.2-12 T. Later on, Oka {\it et al.} showed that such $T^{3}$ behavior reported in the previous works is a consequence of impurity scattering rather than an intrinsic one~\cite{LaFeAsO_Oka}.

Oka {\it et al.} reported the zero field $^{75}$As NQR measurements of $1/T_{1}$ in LaFeAsO$_{1-x}$F$_{x}$~\cite{LaFeAsO_Oka}. NQR has several advantages over NMR, for NQR can avoid the residual DOS induced by vortex. Besides, $^{75}$As nucleus has a nuclear spin $I$ = 3/2 and the recovery curve of the nuclear magnetization is single exponential, so it is more straightforward to obtain $T_{1}$ with high accuracy.

\begin{figure}[th]
 \centerline{\psfig{file=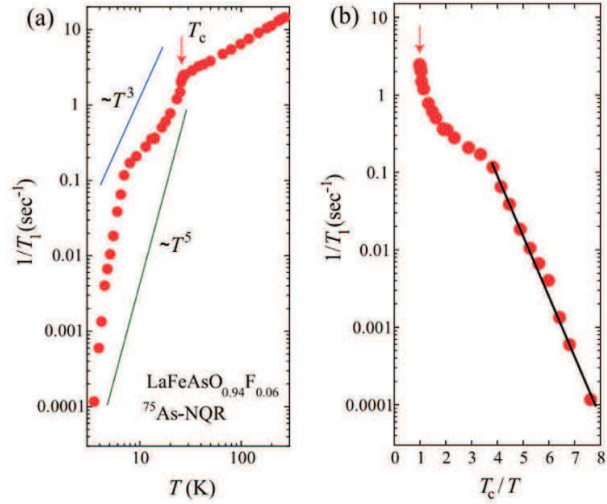,width=8cm}}
 \vspace*{8pt}
 \caption{(color online). (a) The temperature dependence of $1/T_{1}$ for LaFeAsO$_{1-x}$F$_{x}$ ({\it x} = 0.06). (b) Semilogarithmic plot of $1/T_{1}$ vs $T_{c}/T$. The solid line represents the relation ${1}/{{{T}_{1}}}\;\propto \exp \left( {-{{\Delta }_{0}}}/{{{k}_{B}}T}\; \right)$.~\label{f4}}
 \end{figure}

Figure~\ref{f4} shows the temperature dependence of $1/T_{1}$ for $x$ = 0.06. Below $T_{c}$, $1/T_{1}$ decreases steeply due to the opening of the superconducting gaps. The hump structure at $T$ $\sim$ 0.4$T_{c}$ is due to the multiple-gap character as found in PrFeAsO$_{1-x}$F$_{x}$. The $T$ variation at low temperature is much stronger than $T^{3}$ and even stronger than $T^{5}$, as can be clearly seen in Fig.~\ref{f4} (a). In fact, $1/T_{1}$ decreases exponentially below 0.4$T_{c}$. In Fig.~\ref{f4} (b), $1/T_{1}$ is plotted against $T_{c}/T$ in a semilogarithmic scale. As indicated by the solid line, the $1/T_{1}$ below $T$ = 0.4$T_{c}$ clearly follows the relation ${1}/{{{T}_{1}}}\;\propto \exp \left( {-{{\Delta }_{0}}}/{{{k}_{B}}T}\; \right)$. This is clear and direct evidence that the superconducting state is fully gapped.

\begin{figure}[th]
 \centerline{\psfig{file=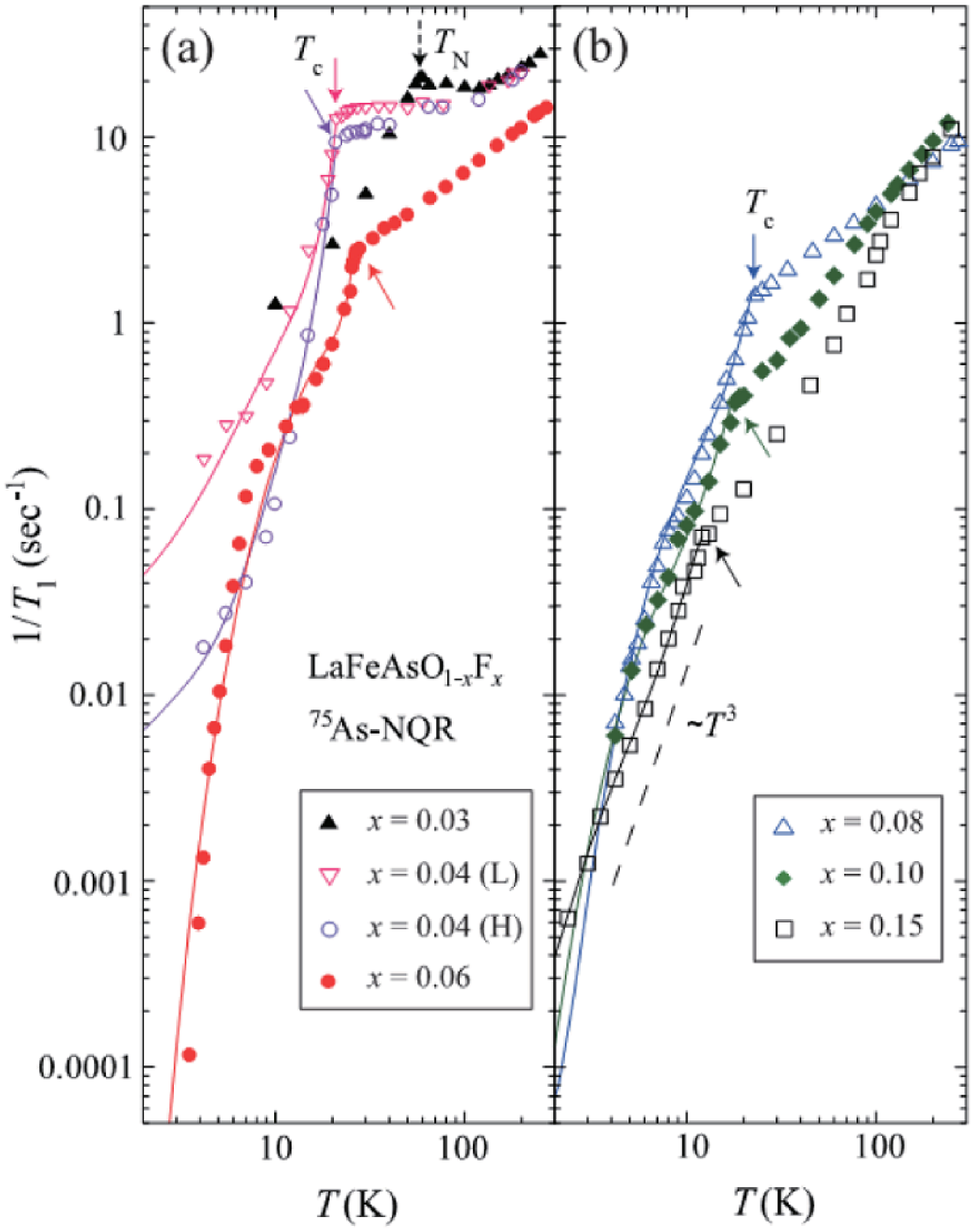,width=8cm}}
 \vspace*{8pt}
 \caption{(color online) The temperature dependence of $1/T_{1}$ for $x$ = 0.03, 0.04, 0.06 (a) and for $x$ = 0.08, 0.10, 0.15 (b). Solid curves below $T_{c}$ for $x$ $\geqslant$ 0.04 are the simulations based on a $s^{\pm}$-wave superconducting gap model with impurity scattering (see the text). The dashed line indicates the relation $1/T_{1} \propto T^{3}$. The dotted and solid arrows indicate $T_{N}$ and $T_{c}$, respectively.~\label{f5}}
 \end{figure}

Figure~\ref{f5} shows the evolution of $1/T_{1}$ with increasing doping level {\it x}. For $x$ = 0.06 - 0.10, $1/T_{1}$ shows a marked hump structure around $T$ $\sim$ 0.4$T_{c}$ and is followed by a still sharper decrease below. However, this behavior of $1/T_{1}$ changes gradually, as to decrease less and less steeply as $x$ increases. Eventually, for $x$ = 0.15, the hump structure disappears completely. Instead, a simple $1/T_{1}$ $\sim$ $T^{3}$ behavior emerges. For $x$ = 0.04, the $T$-dependence of $1/T_{1}$ is much slower.

By assuming $s^{\pm}$-wave symmetry with impurity scattering, the doping evolution of the $T$-dependence of $1/T_{1}$ below $T_{c}$ was reproduced. For $s^{\pm}$-wave symmetry, the gaps fully open but change signs on different Fermi surfaces~\cite{band cal_Kuroki,band cal_Lee dh,band cal_mazin}. The lack of the coherence peak just below $T_{c}$ can be understood within this scenario as due to the sign changed gap and impurity scattering. By introducing the impurity scattering parameter $\eta$ in the energy spectrum in the form of $E = \omega + i\eta$, and using Eq.~\ref{e1}, the $1/T_{1}$ in the superconducting state was well fitted as shown in Fig.~\ref{f5}. A three bands model corresponding to ARPES measurement~\cite{ARPES_ding} was employed, where $N_i$ is the DOS coming from band $i$ ($i$ = 1, 2, 3 denote the $\gamma$, $\beta$, and $\alpha$ bands found in ARPES), $\Delta _{1}^{+}$, $\Delta _{2}^{-}$ and $\Delta _{3}^{-}$ are the gaps on respective Fermi surfaces. The obtained fitting parameters are summarized in Table~\ref{t1}.

It is noticeable that in the $x$ = 0.15 sample, the $1/T_{1} \sim T^{3}$  behavior can be explained as a result of the impurity scattering which brings about a finite DOS. For the $x$ = 0.04 sample where two phases coexist (phase seperation), the weak $T$-dependent $1/T_{1}$ can also be fitted by the same model, with an additional feature that a large $\eta$ is needed to explain the low-$T$ behavior. This can be understood if the two phases coexist at nanoscale, so that one phase acts as an impurity scatterer for the other.

\begin{table}[pt]
\caption{The fitting parameters $\Delta _{1}^{+} = \Delta _{3}^{-}$, $\Delta _{2}^{-}$, $\eta$ in the unit of $k_{B}T_{c}$ and the ratio $N_{1}:N_{2}:N_{3}$}
\label{t1}
\begin{tabular}{ccrrcc}
\hline
   $x$ &      $T_{c}$(K) &       $\Delta _{1}^{+} = \Delta _{3}^{-}$        &            $\Delta _{2}^{-}$&    $\eta$  &   $N_{1}:N_{2}:N_{3}$ \\
\hline
   0.04(L) &         21 &        4.5 &       0.93 &       0.39 & 0.335:0.33:0.335 \\

   0.04(H) &         21 &       4.58 &       1.63 &       0.27 & 0.38:0.24:0.38 \\

      0.06 &         27 &       5.62 &       1.11 &      0.006 & 0.30:0.40:0.30 \\

      0.08 &         23 &       3.37 &       0.92 &       0.03 & 0.303:0.394:0.303 \\

       0.1 &         18 &          3 &       0.83 &      0.035 & 0.305:0.39:0.305 \\

      0.15 &         12 &       2.62 &       0.79 &       0.15 & 0.31:0.38:0.31 \\

Ba$_{0.68}$K$_{0.32}$Fe$_{2}$As$_{2}$ &       38.5 &        4.7 &       0.96 &      0.022 & 0.44:0.12:0.44 \\
\hline
\hline
\end{tabular}
\end{table}

\subsection{Li$_{x}$FeAs}

\begin{figure}[th]
 \centerline{\psfig{file=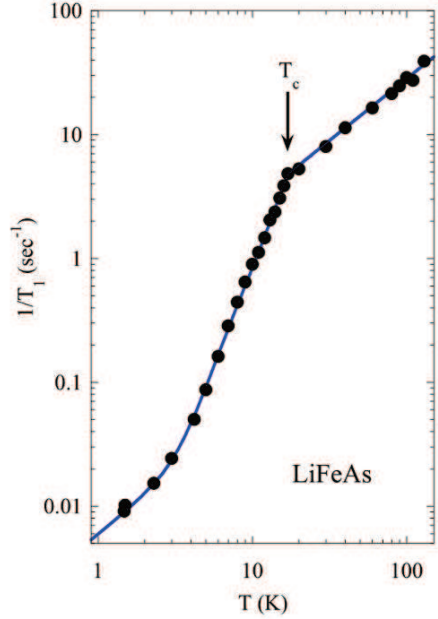,width=6cm}}
 \vspace*{8pt}
 \caption{(Color online) The temperature dependence of $1/T_{1}$ measured by NQR for LiFeAs. The curve
below $T_{c}$ is fit to the $s^{\pm}$-wave model in the presence of impurity scattering. \label{f6}}
 \end{figure}

Li {\it et al.} reported the $^{75}$As NQR studies on polycrystalline Li$_{x}$FeAs samples~\cite{LiFeAs_Li}. They did measurements on two polycrystalline samples of Li$_{x}$FeAs with nominal $x$ = 0.8 and 1.1, and found that the physical properties including the NMR results are the same for the two samples. $T_{c}$ for the nominal $x$ = 0.8 sample is 17 K at zero magnetic field and 16 K at $H$ = 7.3 T. Figure~\ref{f6} shows the temperature dependence of the spin-lattice relaxation rate $1/T_{1}$. The $1/T_{1}$ decreases below $T_{c}$ without a coherence peak as in the 1111 compounds. with further decreasing temperature, $1/T_{1}$ becomes to be proportional to $T$ below $T \sim T_{c}/4$, which indicates that a finite DOS is present. The results can also be understood by assuming $s^{\pm}$-wave symmetry with impurity scattering. In this work, the authors used a two-gap $s^{\pm}$-wave model to fit the data as shown in Fig.~\ref{f6}. The obtained parameters are $\Delta _{1}^{+} = 3.0 k_{B}T_{c}$, $\Delta _{2}^{-} = 1.3 k_{B}T_{c}$, $N_{1}:N_{2} = 0.5:0.5$ and $\eta = 0.26k_{B}T_{c}$.

\subsection{Ba$_{1-x}$K$_{x}$Fe$_{2}$As$_{2}$}

\begin{figure}[th]
 \centerline{\psfig{file=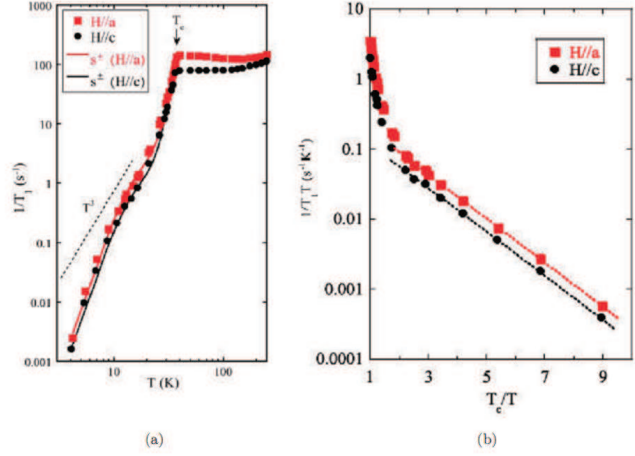,width=9cm}}
 \vspace*{8pt}
 \caption{(color online) The temperature dependence of $1/T_{1}$ for single-crystal Ba$_{0.68}$K$_{0.32}$Fe$_{2}$As$_{2}$. (a) The dashed line shows the $T^{3}$ variation. The curves below $T_{c}$ are fitted to a two-gap $s^{\pm}$ model using the same parameters for both directions. (b) The semilog plot of $1/T_{1}T$ vs $T_{c}/T$. \label{f7}}
 \end{figure}

\begin{figure}[th]
\centerline{\psfig{file=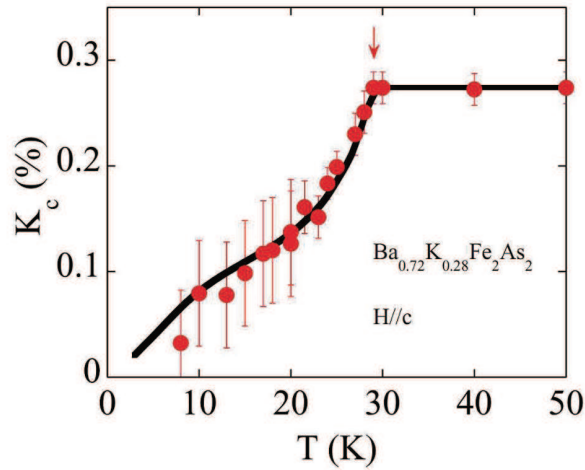,width=8cm}}
\vspace*{8pt}
\caption{(color online) The Knight shift of Ba$_{0.72}$K$_{0.28}$Fe$_{2}$As$_{2}$ with $H$~$\|$~$c$-axis. The arrow indicates $T_{c}$. The curve below $T_{c}$ is fitted to a two-gap model. \label{f8}}
\end{figure}

BaFe$_{2}$As$_{2}$ belongs to the ThCr$_{2}$Si$_{2}$ structure, which is the same as the famous heavy fermion superconductor CeCu$_{2}$Si$_{2}$~\cite{CeCu2Si2}. By substitution K for Ba~\cite{Ba122_38K}, or P for As~\cite{122_P}, or Co/Ni for Fe~\cite{122_Co,122_Ni}, one can get a hole-doped, isovalent-doped, or electron-doped superconductor, respectively. In the copper-oxide high-$T_{c}$ superconductors, there are many differences between hole doping and electron doping. For example, hole-doped compounds have a strong electron correlation in under-doped and optimal-doped region~\cite{zheng_PRL}, while for the electron-doped ones the correlation is weak~\cite{zheng_PRL2}. Therefore, BaFe$_{2}$As$_{2}$ provides an opportunity for studying similarities and  dissimilarities between hole doping and electron doping. In addition, the 122 structure has two Fe-As layers per unit cell but 1111 has only one, so it also provides a good opportunity to study the relationship between structure and superconductivity.

Li {\it et al.} reported the $^{75}$As NMR results in a high quality single-crystal Ba$_{0.68}$K$_{0.32}$Fe$_{2}$As$_{2}$ ($T_{c}$ = 38.5 K) grown by the self-flux method~\cite{Ba122_li}. The $T_{c}$ is 37.6 K for $H$ (=7.5 T) $\parallel$ $a$ axis and 36.4 K for $H$ (=7.5 T) $\parallel$ $c$ axis. Figure~\ref{f7} (a) shows the temperature dependence of $1/T_{1}$. The $1/T_{1}$ decreases rapidly below $T_{c}$. The variation is much faster than $T^{3}$ at low $T$. The $1/T_{1}$ also shows a hump structure around half $T_{c}$, indicating multiple gaps. To see the low-$T$ behavior more clearly, $1/T_{1}T$ was plotted as a function of $T_{c}/T$ in Fig.~\ref{f7} (b). As can be seen there, $1/T_{1}T$ shows a good exponential behavior below 17 K, which is the same as the case in LaFeAsO$_{1-x}$F$_{x}$. This is strong evidence for fully opened gaps. By using the three band $s^{\pm}$-wave model mentioned in section~\ref{LaFeAsO(F)}, $1/T_{1}$ data can be fitted well as shown in Fig.~\ref{f7}. The fitting parameters obtained are shown in Table~\ref{t1}.

The multiple-gap feature was also found in the Knight shift in this family. Matano {\it et al.} reported the Knight shift measurements on a single crystal Ba$_{0.72}$K$_{0.28}$Fe$_{2}$As$_{2}$ ($T_{c}$ = 31.5 K) by $^{75}$As-NMR~\cite{Ba122_Matano}. Figure~\ref{f8} shows the temperature dependence of  Knight shift with $H$ parallel to the $c$-axis. $K_{c}$ decreases below $T_{c}$ and also shows a step-wise behavior at about $T_{c}$/2, which is quite similar to the case for PrFeAs0$_{0.89}$F$_{0.11}$.

\subsection{Summary on the superconducting state}

\begin{figure}[th]
\centerline{\psfig{file=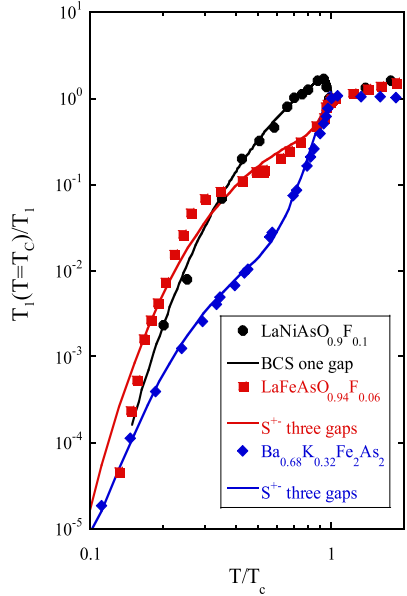,width=6cm}}
\vspace*{8pt}
\caption{(color online) The normalized $T$-dependence of $1/T_{1}$ for typical iron-based superconductors and nickel-analog. The solid curves are simulations by different models. \label{f9}}
\end{figure}

In summary, for iron-based superconductors the temperature dependence of the spin-lattice relaxation rate $1/T_{1}$ under $T_{c}$ is very unique. If one compares with nickel-based LaNiAsO$_{1-x}$F$_{x}$~\cite{Tabuchi_Ni1111} as shown in Fig.~\ref{f9}, the characteristic becomes clear. The result for LaNiAsO$_{1-x}$F$_{x}$ indicates that it is a conventional $s$-wave superconductor with a single fully opened gap. A well-defined coherence peak arises just below $T_{c}$ and $1/T_{1}$ follows an exponential decay. But for iron-based samples, there is no coherence peak, with a hump structure around $T_{c}$/2. The difference may be ascribed to the different topology of the Fermi surfaces. For iron-arsenides, Fermi surfaces are well nested. For LaNiAsO$_{1-x}$F$_{x}$, however, there is no such Fermi surface nesting~\cite{Ni1111_LDA}. These results suggest the importance of the Fermi surface topology in the iron-arsenides.

\section{Spin fluctuations and its relationship with superconductivity}

In cooper oxides, it is widely believed that the antiferromagnetic spin fluctuation(AFSF) is essential to induce high-$T_{c}$ superconductivity. In iron-pnictides, superconductivity also emerges when antiferromagnetic(AFM) order is suppressed via chemical substitution or pressure. In the paramagnetic state, the antiferromagnetic spin fluctuations (AFSF) have been found by NMR/NQR experiments, but whether the AFSF is associated with superconductivity or not was hotly debated.

\begin{figure}[th]
\centerline{\psfig{file=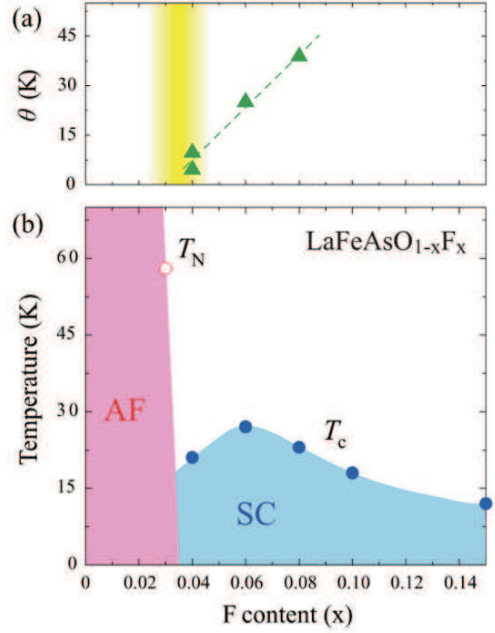,width=7cm}}
\vspace*{8pt}
\caption{(color online) Phase diagram for LaFeAsO$_{1-x}$F$_{x}$ obtained by Oka {\it et al.} AF
and SC denote the antiferromagnetically ordered and superconducting states, respectively. (a) $x$ dependence of $\theta$. The dotted
line is a guide to the eyes. The shade indicates the region of
phase separation. (b) $x$ dependence of $T_{N}$ and $T_{c}$ determined by
NQR measurements. \label{f10}}
\end{figure}

\begin{figure}[th]
\centerline{\psfig{file=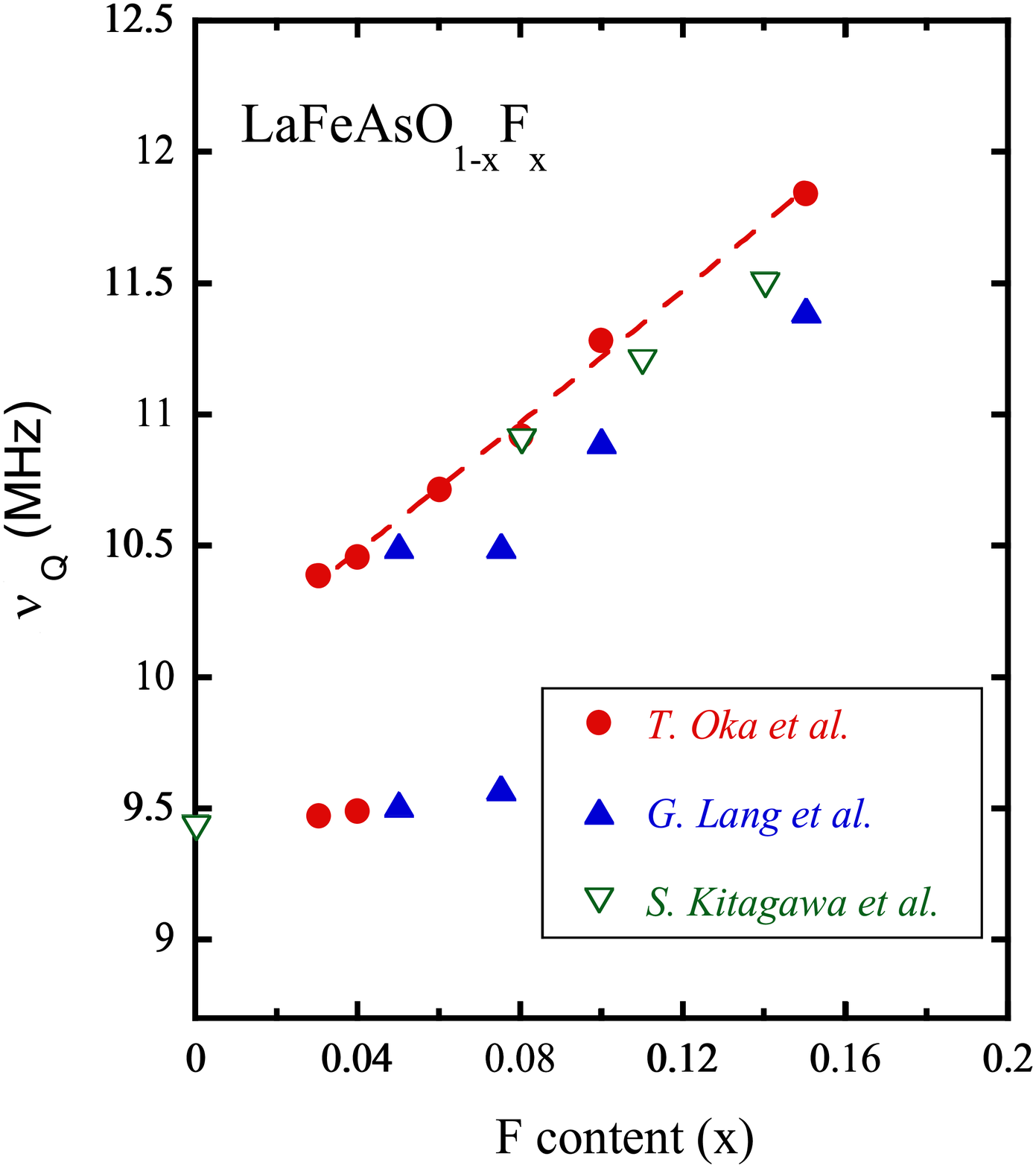,width=7cm}}
\vspace*{8pt}
\caption{(color online) Nominal F-content dependence of $^{75}$As-NQR frequency ($\nu_{Q}$) for LaFeAsO$_{1-x}$F$_{x}$ from different groups. The dotted line is a guide to the eyes.
\label{f11}}
\end{figure}

For 1111 type LaFeAsO$_{1-x}$F$_{x}$, Nakai {\it et al.} reported $^{75}$As NMR studies on a series of polycrystalline samples (0 $\leqslant$ $x$ $\leqslant$ 0.14)~\cite{LaFeAsO_nakai_AFSF}. Their results showed that AFSF are present for $x$ = 0 and 0.04 but are strongly suppressed by electron doping. For $x$ = 0.11 where $T_{c}$ is maximum, no AFSF is observed. $1/T_{1}T$ significantly depends on F doping level, while $T_{c}$ was almost unchanged from $x$ = 0.04 to 0.11. From this observation, they suggested that AFSF may not play an important role in superconductivity. Mukuda {\it et al.} drew the
similar conclusion~\cite{LaFeAsO_Mukuda}.

 Oka {\it et al.} reported a systematic $^{75}$As NQR study of LaFeAsO$_{1-x}$F$_{x}$ with $x$ = 0.03, 0.04, 0.06, 0.08, 0.10, and 0.15~\cite{LaFeAsO_Oka}. $T_{c}$ was determined by ac susceptibility measurements using the in-situ NQR coil and by $1/T_{1}$. The results by the two methods agree well. They found a relationship between the AFSF and superconductivity which is quite different from the previous reports.

\begin{figure}[th]
\centerline{\psfig{file=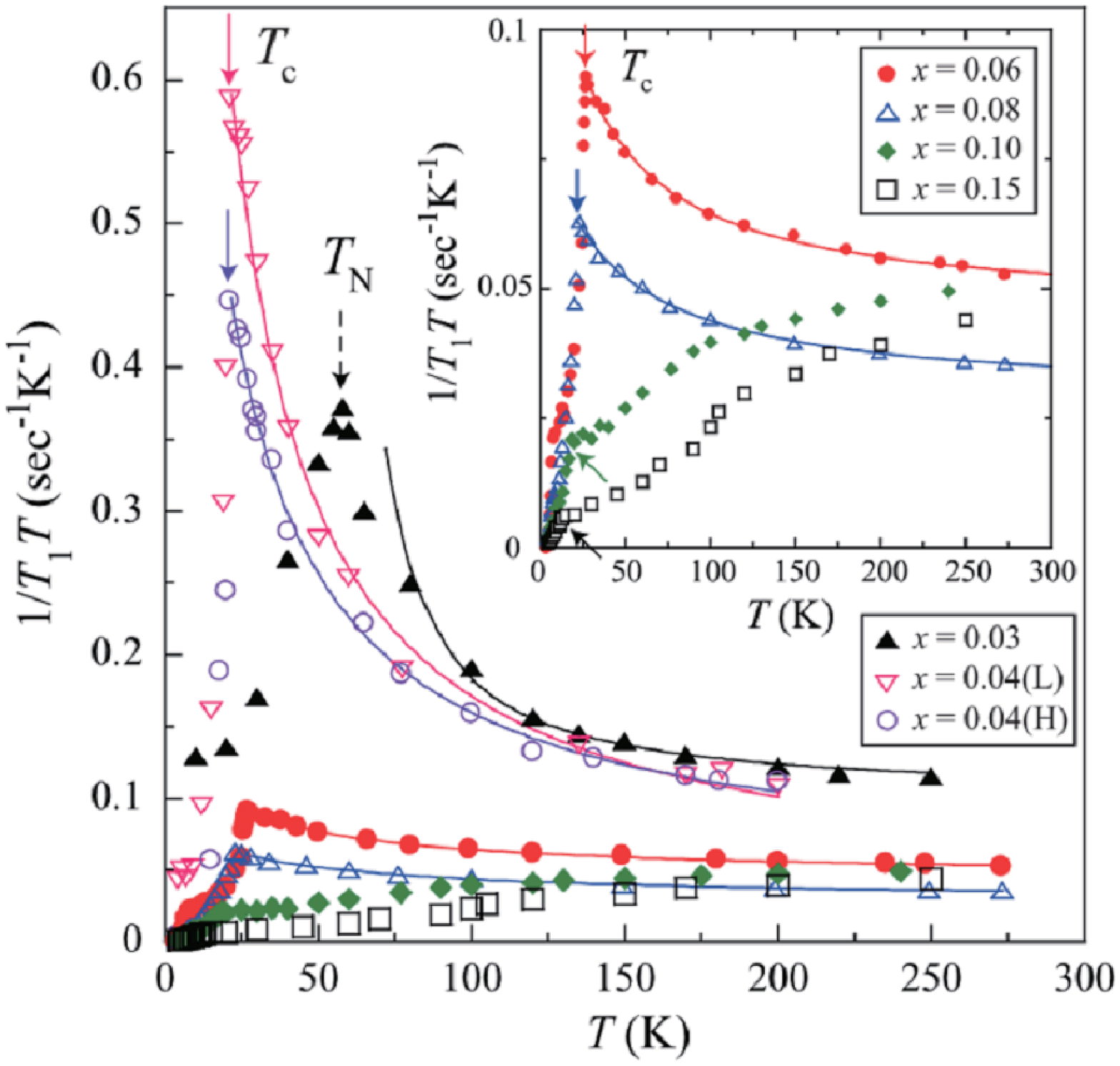,width=8cm}}
\vspace*{8pt}
\caption{ (color online) The temperature dependence of $1/T_{1}T$ for LaFeAsO$_{1-x}$F$_{x}$. The curves above $T_{N}$ or $T_{c}$ are fitted to the AFSF theory (see the text). The inset is the enlarged part for 0.06 $\leqslant$ $x$ $\leqslant$  0.15. \label{f12}}
\end{figure}

Figure~\ref{f10} shows the phase diagram for LaFeAsO$_{1-x}$F$_{x}$ obtained by Oka {\it et al.} For $x$ = 0.03, an antiferromagnetic order with $T_{N}$ = 58 K is found, while no superconductivity is observed. Bulk superconductivity with $T_{c}$ = 21 K arises for $x$ = 0.04, and the highest $T_{c}$ = 27 K is found in the low-doping regime at $x$ = 0.06. The most important feature of this phase diagram is that the superconducting region has a dome-like shape against $x$. In the earlier reports, $T_{c}$ forms a plateau for a wide range 0.04 $\leqslant$ $x$ $\leqslant$ 0.11.~\cite{LaFeAsO_nakai_AFSF} The failure of obtaining higher $T_{c}$ in the low-doping regime in the earlier works is probably due to sample inhomogeneity as evidenced by the broader (in fact, two-peak-featured) NQR spectrum. Also, $T_{c}$ was determined by resistivity measurement which gives a higher value than determined by susceptibility. Furthermore, the nominal $x$ value was probably larger than the actual F content in the compound. Oka {\it et al.} suggested that $\nu_{Q}$ should be used as a tool to determine the real doping level. Figure~\ref{f11} shows the nominal F-content dependence of $^{75}$As-NQR frequency ($\nu_{Q}$) for LaFeAsO$_{1-x}$F$_{x}$ from different groups. Judging from the results, it was pointed out that the nominal $x$ $\sim$ 0.06 of Lang {\it et al.}~\cite{LaFeAsO_lang} actually has a real $x$ close to $x$ $\sim$ 0.04 of Oka {\it et al}. Similarly, the nominal $x$ $\sim$ 0.11 of Kitagawa {\it et al}~\cite{LaFeAsO_S Kawasaki} would have a real $x$ close to $x$ $\sim$ 0.09 of Oka {\it et al.}.

 Figure~\ref{f12} shows the temperature dependence of the $1/T_{1}T$ in LaFeAsO$_{1-x}$F$_{x}$. None of the samples shows a Korringa relation $1/T_{1}T$ = const expected for a conventional metal. Above $T_{N}$ of $x$ = 0.03, $1/T_{1}T$ increases with decreasing $T$ due to the AFSF. Such AFSF persists in $x$ = 0.04, 0.06, and 0.08, where $1/T_{1}T$ increases with decreasing $T$ down to $T_{c}$.

 In general, $1/T_{1}$ is related to the transverse fluctuating hyperfine fields and can be written as~\cite{moriya}

\begin{equation}\label{e2}
\frac{1}{{{T}_{1}}}=\frac{{{\gamma }^{2}}}{2}\int_{-\infty }^{\infty }{\text{d}t}\cos \left( {{\omega }_{0}}t \right)\left\langle \delta {{H}_{+}}\left( t \right)\delta {{H}_{-}}\left( 0 \right) \right\rangle
\end{equation}

where $<>$ denotes the statistical average, $\delta H$ is connected to the Fe moment $S$ by $\delta H = A\cdot S$, where $A$ is the $q$-dependent hyperfine coupling tensor between the As nucleus and Fe spins. One therefore obtains

\begin{equation}\label{e3}
\frac{1}{{{T}_{1}}}=\frac{{{\gamma }^{2}}}{2}\sum\limits_{q}{{{A}_{q}}{{A}_{-q}}\int_{-\infty }^{\infty }{\text{d}t}\cos \left( {{\omega }_{0}}t \right)\left\langle S_{q}^{+}\left( t \right)S_{-q}^{-}\left( 0 \right) \right\rangle }
\end{equation}

using the fluctuation-dissipation theorem, and considering $\hbar {{\omega }_{0}}<<{{k}_{B}}T$, one obtains

 \begin{equation}\label{e4}
\frac{1}{{{{T_1}T}}} = \frac{{2{\gamma ^2}{k_B}}}{{{{\left( {{\gamma _e}\hbar } \right)}^2}}}\sum\limits_q {{A_q}{A_{ - q}}{\textstyle{{{{\chi ''}_ \bot }\left( {q,{\omega _0}} \right)} \over {{\omega _0}}}}}
\end{equation}

 so $1/T_{1}T$ is related to the low-energy dynamical susceptibility ${{{\chi }''}_{\bot }}$.

Oka {\it et al.} analyzed their results by assuming that $1/T_{1}T$ comes from two contributions,

\begin{equation}\label{e5}
\frac{1}{{{T_1}T}} = {\left( {\frac{1}{{{T_1}T}}} \right)_{AF}} + {\left( {\frac{1}{{{T_1}T}}} \right)_0} = \frac{C}{{T + \theta }} + {\left( {\frac{1}{{{T_1}T}}} \right)_0}
\end{equation}

Here, the first term described the contribution from the antiferromagnetic wave vector, and the second term is the contribution from $q = 0$, namely, the DOS at the Fermi level. For the first part, the theory for weakly antiferromagnetic metals was employed, which yields a Curie-Weiss $T$-dependence of $1/T_{1}T$. The fitting curve is shown in Fig.~\ref{f12}. For $x$ = 0.03, $\theta$ is simply -$T_{N}$. For $x$ = 0.04, 0.06, and 0.08, $\theta$ is 10, 25, and 39 K, respectively. The increase of $\theta$ with increasing $x$ means that the system moves away from the magnetic instability (MI) where $\theta$ = 0 K. With further doping, for $x$ = 0.10 and 0.15, no AFSF is seen. Instead, $1/T_{1}T$ decreases with decreasing $T$, which was recently explained by the loss of the DOS due to a topological change of the Fermi surface.~\cite{ikeda}

\begin{figure}[th]
\centerline{\psfig{file=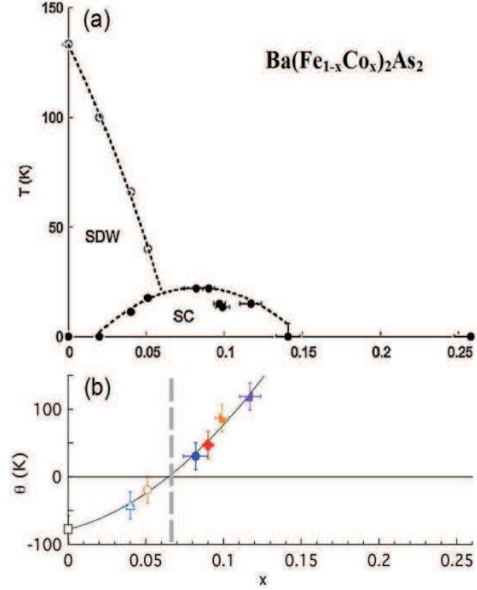,width=7cm}}
\vspace*{8pt}
\caption{ (color online) (a) Phase diagram for Ba(Fe$_{1-x}$Co$_{x}$)$_{2}$As$_{2}$ obtained by Ning {\it et al.} (b) The concentration dependence of $\theta$. Solid curves are guides for the eyes. \label{f13}}
\end{figure}

The remarkable finding is that the highest $T_{c}$ = 27 K is
realized at $x$ = 0.06, which is away from the MI. This
situation is quite similar to La$_{2-x}$Sr$_{x}$CuO$_{4}$~\cite{La214}. In the scenario of spin fluctuation-mediated superconductivity, this can be understood as follows. At high doping levels, the decrease of $T_{c}$ is due to the weakening of the AFSF. In the vicinity of the MI, on the other hand, the too strong low-energy fluctuation acts as pair breaking~\cite{pair breaking}. Therefore, a maximal $T_{c}$ is realized at some point
away from the MI with moderate AFSF.

Measurements on 122 type iron-pnictides have also suggested that the AFSF correlates with superconductivity. Ning {\it et al.} investigated the spin dynamics in the single crystal Ba(Fe$_{1-x}$Co$_{x}$)$_{2}$As$_{2}$ by $^{75}$As NMR~\cite{Ning_BaCo}. They found that AFSF detected by $1/T_{1}T$ exists in  almost the entire doping range. The phase diagram they obtained is quite similar to that obtained by Oka {\it et al.}, as shown in Fig.~\ref{f13}.

\section{Spin-fluctuations anisotropy and the spin-orbit coupling}

What is the difference between the spin fluctuations observed in iron-pnictides and those in copper oxides?

\begin{figure}[th]
\centerline{\psfig{file=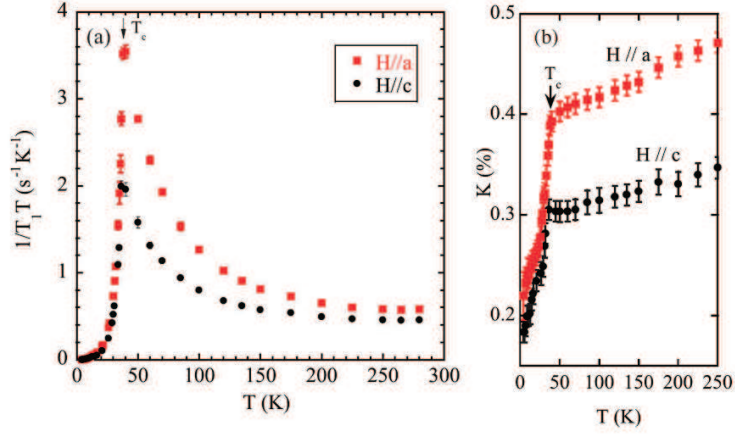,width=10cm}}
\vspace*{8pt}
\caption{(color online) The temperature dependence of $1/T_{1}T$ (a) and Knight shift (b) for Ba$_{0.68}$K$_{0.32}$Fe$_{2}$As$_{2}$, respectively. The arrow indicates $T_{c}$.
\label{f14}}
\end{figure}

\begin{figure}[th]
\centerline{\psfig{file=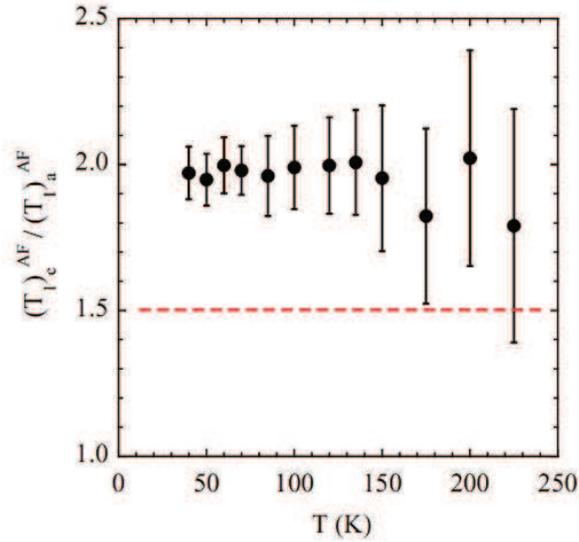,width=8cm}}
\vspace*{8pt}
\caption{ (Color online) The temperature dependence of the anisotropy of $T_{1}$ due
to AFSF in Ba$_{0.68}$K$_{0.32}$Fe$_{2}$As$_{2}$. The dashed line marks the value for isotropic AFSF.
\label{f15}}
\end{figure}

Li {\it et al.} reported $^{75}$As NMR studies on the optimally doped single-crystal Ba$_{0.68}$K$_{0.32}$Fe$_{2}$As$_{2}$ with $T_{c}$ = 38.5 K~\cite{Ba122_li}. Figure~\ref{f14} (a) and (b) show the $T$ dependence of $1/T_{1}T$ and Knight shift respectively. The $1/T_{1}T$ increases with decreasing $T$ down to $T_{c}$, which indicates strong AFSF. By using the theory described by Eq.~\ref{e5}, they assume $\frac{1}{{{T_1}T}} = {\left( {\frac{1}{{{T_1}T}}} \right)_{AF}} + {\left( {\frac{1}{{{T_1}T}}} \right)_0} = \frac{C}{{T + \theta }} + {\left( {\frac{1}{{{T_1}T}}} \right)_0}$, where ($1/T_{1}T$)$_{AF}$ is due to the susceptibility at the AF wave vector Q, and ($1/T_{1}T$)$_{0}$ is due to $s$-band electrons and the orbital hyperfine interaction. By taking the averaged value of $1/T_{1}T$ at $T$ = 250 K as ($1/T_{1}T$)$_{0}$, ($1/T_{1}T$)$_{AF}$ is then obtained. Figure~\ref{f15} shows the ratio of $1/T_{1}T$ due to AFSF, $(T_{1})^{AF}_{c}/(T_{1})^{AF}_{a}$, whose value is about 2. This result indicates that the AFSF is anisotropic in the spin space, as elaborated below.

For Q = ($\pi$,0) and Q = (0,$\pi$) AFSF, one has~\cite{kitagawa_HF}
\begin{equation}\label{e6}
{\bf{A}}\left( {\pi ,0} \right) = \left( {\begin{array}{*{20}{c}}
0&0&A\\
0&0&0\\
A&0&0
\end{array}} \right)
,
{\bf{A}}\left( {0,\pi } \right) = \left( {\begin{array}{*{20}{c}}
0&0&0\\
0&0&A\\
0&A&0
\end{array}} \right)
\end{equation}

using Eq.~\ref{e4}, one then obtains
\begin{equation}\label{e7}
{R_{AF}} = \frac{{\left( {1/{T_1}} \right)_a^{AF}}}{{\left( {1/{T_1}} \right)_c^{AF}}} = \frac{{{{\chi ''}_a}\left( {{\omega _0},Q} \right) + {{\chi ''}_b}\left( {{\omega _0},Q} \right)}}{{2{{\chi ''}_c}\left( {{\omega _0},Q} \right)}} + \frac{1}{2}
\end{equation}

\begin{figure}[th]
\centerline{\psfig{file=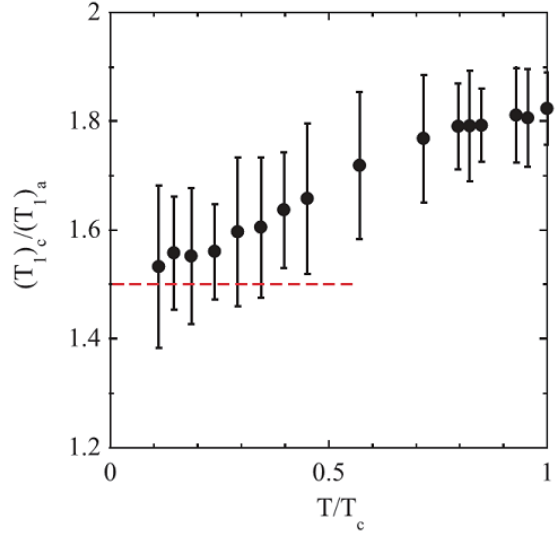,width=8cm}}
\vspace*{8pt}
\caption{ (Color online) The temperature dependence of the $T_{1}$ anisotropy below
$T_{c}$ for Ba$_{0.68}$K$_{0.32}$Fe$_{2}$As$_{2}$. The dashed straight line indicates the value for isotropic AFSF. \label{f16}}
\end{figure}

If ${\chi ''_a}\left( {{\omega _0},Q} \right) = {\chi ''_b}\left( {{\omega _0},Q} \right) = {\chi ''_c}\left( {{\omega _0},Q} \right)$, namely, if the AFSF is isotropic in the spin space, then $R_{AF}$ = 1.5. The observed $R_{AF}$ shown in Fig.~\ref{f15} is much larger than 1.5, which follows from Eq.~\ref{e7} that ${{{\chi }''}_{a}}\left( {{\omega }_{0}},Q \right)$ (${{{\chi }''}_{b}}\left( {{\omega }_{0}},Q \right)$) is larger than ${{{\chi }''}_{c}}\left( {{\omega }_{0}},Q \right)$ by about 50\%.

They proposed that the anisotropy of AFSF results from spin-orbit coupling (SOC) that mixes spin and orbital freedoms so that the magnetic susceptibility bears some orbital character, which is anisotropic. They adopted a two-band model involving $d_{xz}$ and $d_{yz}$ to calculate the anisotropy and obtained $R_{AF}$ $\sim$ 2 with reasonable parameters~\cite{Ba122_li}.

An interesting feature is that the ratio $R$ decreases below $T_{c}$ and approaches the isotropic AFSF value 1.5 as shown in Fig.~\ref{f16}. Below $T_{c}$, it is less trivial to subtract the contribution of ($1/T_{1}T$)$_{0}$, so they simply plot the ratio of the observed $(T_{1})_{c}/(T_{1})_{a}$. The decrease of $(T_{1})_{c}/(T_{1})_{a}$ to 1.5 at $T$ $\sim$ 0 implies that the gaps are fully opened and all the electrons are paired so that SOC effect vanishes. This is another evidence for nodeless gap and implies that the AFSF persists in the superconducting state.

The anisotropy of Fe spin fluctuations in the spin space, i.e., ${{{\chi }''}_{\pm }}\left( {{\omega }_{0}},Q \right) > {{{\chi }''}_{z}}\left( {{\omega }_{0}},Q \right)$ (where $z$ is along the $c$ direction), is a noteworthy phenomenon. This feature is similar to that in cobalt oxide~\cite{NaCoO_matano}, but contrary to that in copper oxides where the AFSF is isotropic. In the past 20 years, the spin fluctuation mediated superconducting mechanism has studied intensively. In this scenario, $T_{c}$ is related to the characteristic energy scale of spin fluctuation and the coherence length. The greater the energy scale and the coherence length, the higher the $T_{c}$~\cite{AFSF_Tc_moriya,AFSF_Tc_Morthoux,AFSF_Tc_zheng}. However, as far as we know, few studies are focused on relationship between the anisotropy of AFSF and superconductivity. In a 3D Hubburd model, Scalapino {\it et al.} found that the anisotropy of the spin fluctuations is detrimental to superconductivity~\cite{Hubburd}. Perhaps this is one of the reasons why the $T_{c}$ of the iron-based superconductors is lower than that of the copper oxides.

\section{Summary}

This article reviewed the superconducting and normal state properties of the iron-based high-temperature superconductors  seen via NMR/NQR. The superconducting pairing is in spin-singlet state, and the multiple gaps  originate from the multiple-band structure is an important feature of this class of materials. The antiferromagnetic spin fluctuations found in the normal state are closely associated with the superconductivity, and probably indispensable to high-$T_{c}$ superconductivity. The antiferromagnetic spin fluctuation in the spin space
is anisotropic, which is different from the situation in copper oxide superconductors.
This anisotropy originated from the spin-orbit coupling and is a reflection of the multiple bands structure in this class of new materials.

This article reviewed the superconducting and normal state properties of the iron-based high-temperature superconductors seen via NMR/NQR. The electron pairs in the superconducting state are in the spin-singlet state with multiple nodeless gaps. The multiple gaps originate from the multiple-band structure and is an important feature of this class of materials. The antiferromagnetic spin fluctuations found in the normal state are closely correlated with the superconductivity, and probably indispensable to the high-$T_{c}$ superconductivity. The antiferromagnetic spin fluctuation
is anisotropic in the spin space, which is different from the situation in copper oxide
high temperature superconductors. This anisotropy originated from the spin-orbit
coupling and is a reflection of the multiple-bands structure of this class of new
materials.

\section*{Acknowledgments}

This work was supported by the Ministry of Science and Technology of China (National Basic Research Program Nos. 2011CBA00109 and 2012CB821402).

\end{document}